\documentclass[twocolumn,aps,pre,amsmath,reprint,longbibliography,nofootinbib,superscriptaddress]{revtex4-2}
\usepackage{amsmath,amssymb}
\usepackage{graphicx}
\usepackage{amsthm}
\usepackage{color}
\usepackage[colorlinks=true, bookmarks=true,bookmarksnumbered=true, bookmarkstype=toc, linkcolor=magenta,urlcolor=cyan, citecolor=cyan]{hyperref}
\begin{document}
\title{Exact mapping of a spin glass with correlated disorder to the pure Ising model}
\author{Hidetoshi Nishimori}
\affiliation{Institute of Integrated Research, Institute of Science Tokyo, Nagatsuta-cho, Midori-ku, Yokohama 226-8503, Japan}
\affiliation{Graduate School of Information Sciences, Tohoku University, Sendai 980-8579,
Japan}
\affiliation{RIKEN Interdisciplinary Theoretical and Mathematical Sciences (iTHEMS),
Wako, Saitama 351-0198, Japan}

\date{\today}

\begin{abstract}
We introduce an Ising spin-glass model with correlated disorder which continuously interpolates between the pure ferromagnetic Ising model and the Edwards-Anderson model with symmetric disorder. For this model, we prove that a Nishimori line (NL) can be defined, analogously to the Edwards-Anderson model, on which physical quantities  can be expressed exactly in terms of those of the pure Ising model at a well-defined effective temperature on any lattice in any dimension. For example, the energy on the NL is equal to the energy of the pure Ising model at the effective temperature up to a constant and a trivial factor. More remarkably, the specific heat on the NL equals the energy, not the specific heat, of the pure Ising model at the effective temperature, again up to a constant and a trivial factor. Gauge-noninvariant quantities such as the magnetization and correlation functions are exactly equal to the corresponding quantities of the pure Ising model at the effective temperature. These exact relations imply that the leading critical behavior at that multicritical point for the disorder-correlated model is pure-Ising-like, in contrast to the conventional multicritical universality class of the standard Edwards-Anderson model. Our results motivate further investigations of the relatively unexplored topic of correlations in disorder in spin glasses and related problems.
\end{abstract}
\maketitle

\section{Introduction}
The study of spin glasses has been one of the central topics in the statistical mechanics of disordered systems for several decades \cite{Charbonneau2023,Nishimori2001,Mydosh1993}. Since the seminal work of Edwards and Anderson \cite{Edwards1975}, most theoretical investigations have assumed that disorder variables (i.e., exchange interactions) are independent and identically distributed. While this simplification has led to significant progress, including the celebrated Sherrington-Kirkpatrick model \cite{Sherrington1975} and its Parisi solution via replica symmetry breaking \cite{Parisi1980}, it does not always reflect the complexities of realistic systems. In many physical contexts, correlations between disorder variables are inevitably present due to thermal history, structural constraints, or underlying microscopic interactions. Understanding the effects of such correlations remains a fundamental challenge in disordered systems \cite{Hoyos2011,Bonzom2013,Cavaliere-2019,Munster2021,Nishimori2022,Nishimori2024,Nishimori2025,Nishimori2025tc}.

In this paper, we introduce and study an Ising spin glass model with a type of correlated disorder different from that discussed in our previous papers \cite{Nishimori2024,Nishimori2025,Nishimori2025tc}. The correlation is introduced via the partition function of the system itself, which appears in the numerator of the probability distribution of the disorder variables, in contrast to the constructions in our previous papers, where the partition function appears in the denominator.
This construction allows us to interpolate smoothly between the pure ferromagnetic Ising model (to be called the pure Ising model) and the standard Edwards-Anderson model with symmetric disorder. Correlated disorder is of broad interest not only in magnetic materials but also in quantum information processing. Correlations in noise and errors are known to play a crucial role in the stability and performance of quantum systems \cite{Klesse2005,Clemens2004,Aharonov2008,Preskill2013,Wilen2021}, and the present study may offer broader insights into such phenomena.

A central role in our analysis is played by the Nishimori line (NL) \cite{Nishimori1980,Nishimori1981,Nishimori2001}, a special subspace in the phase diagram where gauge symmetry enables the derivation of various exact results. We prove that numerous physical quantities on the NL of our model can be expressed exactly in terms of the corresponding quantities of the pure Ising model with an effective temperature. In particular, the energy on the NL equals the energy of the pure Ising model with an effective temperature, up to a constant and a trivial factor, implying that the type of the singularity in the energy is shared by the disordered and pure models. An additional remarkable result concerns the specific heat. It is well established that the specific heat of the Edwards-Anderson model on the NL remains finite \cite{Nishimori1981,Nishimori2001}, in contrast to the divergence expected in the pure model at the critical point \cite{Nishimori2010book}. Here we show that the specific heat of the present model is equal to the internal energy, not the specific heat, of the pure Ising model with an effective temperature, again up to a constant and a trivial factor. This unusual identity establishes a direct link between the thermodynamics of the correlated disordered system and the energy of its pure counterpart.

\begin{figure}[ht]
\centering
\includegraphics[width=60mm]{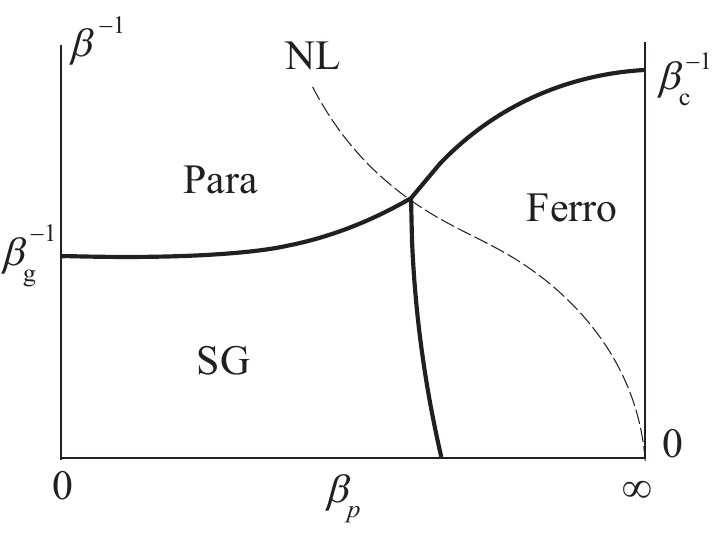}
\caption{
The phase diagram of the model with correlated disorder. $\beta_p$ controls the degree of frustration and the probability of ferromagnetic bonds: $\beta_p \to \infty$ corresponds to the pure ferromagnetic Ising model, and $\beta_p = 0$ corresponds to the Edwards-Anderson model. $\beta^{-1}$ is the temperature, with $\beta_{\rm c}^{-1}$ denoting the critical temperature of the pure Ising model and $\beta_{\rm g}^{-1}$ the spin-glass transition temperature of the Edwards-Anderson model. NL denotes the Nishimori line $\beta=\beta_p$. It is believed that there is no spin-glass phase in two dimensions. The structure of the phase diagram is qualitatively similar to that of the Edwards-Anderson model.
}
\label{fig:phase_diagram}
\end{figure}

Furthermore, correlation functions, including the magnetization as the one-point correlator, are exactly equal to the corresponding quantities of the pure Ising model with an effective temperature.
In addition, several nontrivial identities and an inequality can be derived for correlation functions, which take the same form as in the Edwards-Anderson model. These include the equality between the magnetization and the spin-glass order parameter on the NL, as well as an inequality showing that no ferromagnetic phase exists below the intersection of the NL with the phase boundary. These results coincide with those for the Edwards-Anderson model, suggesting that the NL intersects the phase boundary at a multicritical point, as in the Edwards-Anderson model \cite{LeDoussal1988}. We therefore expect that the structure of the phase diagram is qualitatively similar to that of the Edwards-Anderson model. See Fig.~\ref{fig:phase_diagram}.

These results suggest that the critical behavior at the NL intersection with the phase boundary, which is likely to be the multicritical point, differs from the conventional universality class associated with the multicritical point on the NL in standard Ising spin-glass models~\cite{Honecker2001,LeDoussal1988,Hasenbusch2007,Hasenbusch2008,ParisenToldin2010}. In particular, because correlation functions on the NL reduce exactly to those of the pure Ising model, the leading critical exponents at this critical point are pure-Ising-like. The exact solvability on the NL, in the sense of its direct mapping to the pure Ising model, opens an avenue for rigorously characterizing multicritical phenomena in disordered systems, complementing existing numerical and renormalization-group approaches.

The rest of this paper is organized as follows. Section~\ref{sec:Probability_distr} defines the correlated disorder model and analyzes the properties of correlations. Sections~\ref{sec:E_and_C} and \ref{sec:gauge_non_inv} derive exact results for gauge-invariant and gauge-noninvariant quantities on the NL, respectively. The Gaussian version of the model is briefly described in Sec.~\ref{sec:Gaussian}, and a summary with discussion is given in Sec.~\ref{sec:summary}.

\section{Probability distribution of correlated disorder}
\label{sec:Probability_distr}
In this section we define the model and the probability distribution of disorder variables, and analyze basic statistical properties of the resulting correlated disorder.

The model is defined by the Ising spin-glass Hamiltonian
\begin{align}
H=-\sum_{\langle ij\rangle}\tau_{ij}S_iS_j,
\label{eq:Hamiltonian}
\end{align}
where $\tau_{ij}=\pm 1$ is a quenched disorder variable, $S_i=\pm 1$ is an Ising spin, and the summation runs over all interacting pairs. The lattice structure and spatial dimensionality are arbitrary. The range of interactions is also arbitrary.
Both the Hamiltonian and the inverse temperature $\beta$ are taken to be dimensionless.

Below, we introduce correlations among $\tau\equiv \{\tau_{ij}\}$ through their probability distribution, defined in terms of the partition function of the same spin system. This construction interpolates continuously between the pure Ising model and the Edwards-Anderson model with i.i.d.\ symmetric disorder, while allowing an exact analysis on the NL.

\subsection{Probability distribution}
We define the distribution function of $\tau$ by
\begin{align}
P(\tau) = \frac{1}{A}Z_{\tau}(\beta_{p})e^{\beta_{p}\sum_{\langle ij\rangle}\tau_{ij}},
\label{eq:prob_dist}
\end{align}
where $Z_{\tau}(\beta_p)$ is the partition function
\begin{align}
Z_{\tau}(\beta_p)=\sum_S e^{\beta_p \sum_{\langle ij\rangle}\tau_{ij}S_iS_j}
\end{align}
with a control parameter $\beta_p$, and $A$ is the normalization factor, which depends on $\beta_p$.

To evaluate $A$, it is convenient to perform the sum over $\tau_{ij}=\pm1$ bond by bond, which generates an effective ferromagnetic interaction between spins. Summing $\tau_{ij}=\pm1$ over a single bond yields
\begin{align}
\sum_{\tau_{ij}=\pm1} e^{\beta_p \tau_{ij}(S_iS_j+1)} = 2\cosh(\beta_p(1+S_iS_j)),
\end{align}
so that
\begin{align}
A &= \sum_{\tau}Z_{\tau}(\beta_{p})e^{\beta_{p}\sum_{\langle ij\rangle}\tau_{ij}} \nonumber \\
&=\sum_S \prod_{\langle ij \rangle}(e^{\beta_p+\beta_p S_i S_j}+e^{-\beta_p-\beta_p S_i S_j})\nonumber\\
&= (2\cosh^{2}\beta_{p})^{N_{\rm B}}\sum_{S}\prod_{\langle ij\rangle}(1+S_{i}S_{j}\tanh^{2}\beta_{p}) \nonumber \\
&= f_{1}(\beta_{p})Z_{\rm I}(\beta_{e}).
\label{eq:normalization_const_A}
\end{align}
Here, $N_{\rm B}$ is the number of bonds (number of interacting spin pairs), $Z_{\rm I}$ is the partition function of the pure Ising model, and $\beta_e$ is the effective inverse temperature determined by $\beta_p$ via
\begin{align}
\tanh \beta_{e}(\beta_p)=\tanh^2 \beta_{p}.
\label{eq:effective_beta}
\end{align}
Note that $\beta_e(\beta_p)(<\beta_p)$ is a monotonically increasing function of $\beta_p$, with $\beta_e\to 0$ as $\beta_p\to 0$ and $\beta_e\to\infty$ as $\beta_p\to\infty$.
The prefactor $f_1(\beta_p)$ is defined as
\begin{align}
f_{1}(\beta_{p})=\left(\frac{2\cosh^{2}\beta_{p}}{\cosh\beta_{e}}\right)^{N_{\rm B}}.
\label{eq:f1_definition}
\end{align}

The distribution $P(\tau)$ should be contrasted with the correlated-disorder distributions studied previously~\cite{Nishimori2024,Nishimori2025tc},
\begin{align}
P_{\rm 1}(\tau) = \frac{1}{A_1}\frac{e^{\beta_{p}\sum_{\langle ij\rangle}\tau_{ij}}}{Z_{\tau}(\beta_{p})},
\label{eq:prob_dist_old1}
\end{align}
and
\begin{align}
P_{\rm 2}(\tau) = \frac{Z_{\tau}(\gamma)}{A_2}\frac{e^{\beta_{p}\sum_{\langle ij\rangle}\tau_{ij}}}{Z_{\tau}(\beta_{p})}.
\label{eq:prob_dist_old2}
\end{align}
As discussed in~\cite{Nishimori2024,Nishimori2025tc}, the factor $Z_{\tau}(\beta_p)$ in the numerator of Eq.~(\ref{eq:prob_dist}) enhances the probability of disorder configurations with less frustration. The present choice has two additional distinctive features: (i) it provides a continuous interpolation between the pure Ising model ($\beta_p\to\infty$) and the Edwards-Anderson model with symmetric independent disorder ($\beta_p=0$), an interpolation not available for Eqs.~(\ref{eq:prob_dist_old1}) and (\ref{eq:prob_dist_old2}); (ii) on the NL ($\beta_p=\beta$), many observables reduce exactly to those of the pure Ising model at $\beta_e$, enabling an explicit discussion of critical behavior.

The configurational average with respect to $P(\tau)$ will be denoted as $[\cdots]_{\beta_p}$, with the subscript $\beta_p$ sometimes omitted when no confusion is expected.

\subsection{$(1+\epsilon)$-replica theory}
\label{sec:replica}
The probability distribution of Eq.~(\ref{eq:prob_dist}) may be viewed as a $(1+\epsilon)$-replica theory in the following sense.

The replica theory for the Edwards-Anderson model \cite{Nishimori2001} evaluates the configurational average of the $n$-replicated partition function
\begin{align}
\frac{1}{A_0}\sum_{\tau}e^{\beta_p\sum\tau_{ij}}Z_{\tau}(\beta)^n,
\qquad
(A_0=(2\cosh \beta_p)^{N_{\rm B}}),
\label{eq:replica}
\end{align}
and takes the limit $n=\epsilon \to 0$ to compute the configurational average of the free energy,
\begin{align}
\frac{1}{A_0}\sum_{\tau}e^{\beta_p\sum\tau_{ij}}\big(1+\epsilon \ln Z_{\tau}(\beta)+\mathcal{O}(\epsilon^2)\big).
\end{align}
This method is usually applied to the infinite-range version of the Edwards-Anderson model, the Sherrington--Kirkpatrick model, in which the i.i.d.\ $\pm1$ distribution reduces to a Gaussian distribution according to the central limit theorem.

Similarly, if we set $n=1+\epsilon$ in Eq.~(\ref{eq:replica}) and choose $\epsilon\to 0$, we obtain
\begin{align}
\frac{1}{A_0}\sum_{\tau}e^{\beta_p\sum\tau_{ij}}Z_{\tau}(\beta)\big(1+\epsilon \ln Z_{\tau}(\beta)+\mathcal{O}(\epsilon^2)\big).
\end{align}
Under the NL condition $\beta_p=\beta$, the $\mathcal{O}(\epsilon)$ term may be regarded as the configurational average of the free energy with respect to Eq.~(\ref{eq:prob_dist}), up to the normalization factor. Therefore, the theory developed in the following sections may also be considered as a $(1+\epsilon)$-replicated system of the Edwards-Anderson model on the NL, instead of a theory of correlated disorder. An important difference from the conventional application of replica theory is that we analyze finite-dimensional systems whereas replica theory has been predominantly applied to the infinite-range Sherrington-Kirkpatrick model and its variants except for a limited number of cases, e.g., \cite{Hemmen_1982,LeDoussal1988,Marinari2000,Charbonneau2017,Nahum2025,Patil2026}

In a similar vein, the probability distribution $P_1(\tau)$ of Eq.~(\ref{eq:prob_dist_old1}) can be viewed as the $(-1+\epsilon)$-replica theory.

This replica interpretation is provided only as an auxiliary viewpoint for the present disorder distribution. The exact results derived below do not rely on the replica method.

\subsection{Properties of correlated disorder}
Since $P(\tau)$ cannot be decomposed into a product of independent distributions for each bond when $0<\beta_p<\infty$,
\begin{align}
P(\tau)\ne \prod_{\langle ij \rangle} p(\tau_{ij}),
\end{align}
for any single-bond function $p(\tau_{ij})$, $P(\tau)$ represents correlated disorder. It is instructive to quantify the correlations among the disorder variables $\tau_{ij}$.

To evaluate $[\tau_{12}]$ for an arbitrary bond $(12)$, we compute the unnormalized sum by performing the $\tau$ sum before the $S$ sum:
\begin{align}
&\sum_{\tau}\sum_{S} e^{\beta_p \sum_{\langle ij\rangle} \tau_{ij} S_i S_j} e^{\beta_p \sum_{\langle ij\rangle} \tau_{ij}} \tau_{12} \nonumber\\
&= \sum_{S} \Big({\prod_{\langle ij \rangle}}' (e^{\beta_p S_i S_j + \beta_p} + e^{-\beta_p S_i S_j - \beta_p}) \Big)(1 + S_1 S_2) \sinh 2\beta_p \nonumber \\
&= \frac{(2\cosh^2 \beta_p)^{N_{\rm B}}}{(\cosh \beta_e)^{N_{\rm B}}} \sum_{S} e^{\beta_e \sum_{\langle ij\rangle} S_i S_j} \frac{(1+S_1 S_2) \sinh 2\beta_p}{2\cosh 2\beta_p},
\end{align}
where $\prod_{\langle ij \rangle}'$ denotes the product over all bonds except the pair $(12)$. Dividing by $A = f_1(\beta_p)Z_{\rm I}(\beta_e)$, the $f_1$ and $Z_{\rm I}$ factors cancel and we find
\begin{align}
[\tau_{12}] &= \frac{1}{2} \tanh 2\beta_p \left(1 + \langle S_1 S_2 \rangle_{\beta_e}^{\rm I}\right).
\label{eq:tau_12}
\end{align}
Here $\langle S_1 S_2 \rangle_{\beta_e}^{\rm I}$ is the nearest-neighbor correlation function of the pure Ising model at inverse temperature $\beta_e$.
This average $[\tau_{12}]$ has a singularity at the critical point $\beta_e = \beta_c$ of the pure Ising model in two and higher dimensions. Therefore, the disorder distribution itself becomes singular at a finite value of $\beta_p$, because the Ising criticality enters through $Z_{\rm I}(\beta_e)$ in the normalization factor~(\ref{eq:normalization_const_A}). This is in contrast to $P_{1}$ and $P_{2}$ in Eqs.~(\ref{eq:prob_dist_old1}) and (\ref{eq:prob_dist_old2}), where the normalization factors are trivial~\cite{Nishimori2024,Nishimori2025tc}. In other words, the disorder ensemble inherits the Ising criticality through $Z_{\rm I}(\beta_e)$, and the range of disorder correlations becomes long at $\beta_e=\beta_c$ as we show below.

To quantify correlations between distinct disorder variables, we compute the two-point function,
\begin{align}
&\sum_{\tau} \tau_{12} \tau_{34} e^{\beta_p \sum_{\langle ij\rangle} \tau_{ij} S_i S_j} e^{\beta_p \sum_{\langle ij\rangle} \tau_{ij}} \nonumber \\
&= \sum_{S} \left( {\prod_{\langle ij \rangle}}'' (e^{\beta_p S_i S_j + \beta_p} + e^{-\beta_p S_i S_j - \beta_p}) \right) \nonumber\\
&\quad \times (1+S_1 S_2)(1+S_3 S_4) \sinh^2 2\beta_p \nonumber \\
&= \frac{(2 \cosh^2 \beta_p)^{N_{\rm B}}}{(\cosh \beta_e)^{N_{\rm B}}} \frac{(1+S_1 S_2)(1+S_3 S_4) \sinh^2 2\beta_p}{(2\cosh 2\beta_p)^2},
\end{align}
where $\prod''$ excludes bonds $(12)$ and $(34)$. The two-point function is therefore
\begin{align}
[\tau_{12} \tau_{34}] &= \left( \frac{1}{2} \tanh 2\beta_p \right)^2 \Big(1 + \langle S_1 S_2 \rangle_{\beta_e}^{\rm I} \nonumber\\
&\quad + \langle S_3 S_4 \rangle_{\beta_e}^{\rm I} + \langle S_1 S_2 S_3 S_4 \rangle_{\beta_e}^{\rm I}\Big),
\end{align}
and the connected correlation function is
\begin{align}
&[\tau_{12} \tau_{34}] - [\tau_{12}][\tau_{34}]\notag\\
&= \left( \frac{1}{2} \tanh 2\beta_p \right)^2
\Big(\langle S_1 S_2 S_3 S_4 \rangle_{\beta_e}^{\rm I}
- \langle S_1 S_2 \rangle_{\beta_e}^{\rm I} \langle S_3 S_4 \rangle_{\beta_e}^{\rm I}\Big).
\end{align}
From the generic behavior of correlation functions in the pure Ising model, the connected correlation function of the disorder variables decays exponentially with the distance between the two bonds when $\beta_e$ is away from criticality. At the critical point $\beta_e=\beta_c$, this decay becomes algebraic, and the disorder correlations are long-ranged. This singularity in the disorder ensemble is expected to induce singular behavior in physical observables at $\beta_e=\beta_c$, as will be confirmed by the exact results derived in the following sections.

\section{Energy and specific heat}
\label{sec:E_and_C}
We now evaluate physical quantities on the NL, $\beta_p=\beta$. In this section we focus on two basic gauge-invariant quantities, the internal energy and the specific heat, and show that both can be expressed exactly in terms of the energy of the pure Ising model at the effective inverse temperature $\beta_e$.

\subsection{Energy}
The energy is a function of the inverse temperature $\beta$ and the disorder parameter $\beta_p$, denoted by $E(\beta,\beta_p)$. On the NL ($\beta=\beta_p$), the energy is evaluated as follows:
\begin{align}
&E(\beta,\beta) \nonumber\\
&= -\frac{1}{A}\sum_{\tau}Z_{\tau}(\beta_p)e^{\beta_p\sum \tau_{ij}}\nonumber\\
&\hspace{2cm}\cdot
   \frac{\sum_{S} \left( \sum_{\langle ij\rangle} \tau_{ij}S_i S_j \right)
   e^{\beta \sum \tau_{ij} S_i S_j}}{Z_{\tau}(\beta)}
   \Bigg|_{\beta_p=\beta}\nonumber \\
&= -\frac{1}{A} \sum_{S} \sum_{\tau} \left( \sum_{\langle ij\rangle}
   \tau_{ij} S_i S_j \right) e^{\beta \sum_{\langle ij\rangle}
   \tau_{ij} S_i S_j} e^{\beta \sum \tau_{ij}} \nonumber \\
&= -\frac{1}{A} \sum_{S} \sum_{\tau} \left( \sum_{\langle ij\rangle}
   \tau_{ij} \right) e^{\beta \sum \tau_{ij}} e^{\beta \sum \tau_{ij}
   S_i S_j} \nonumber \\
&= - \sum_{\langle ij\rangle}[\tau_{ij}] \nonumber \\
&= \frac{1}{2} \tanh 2\beta \big(E_{\rm I}(\beta_e) - N_{\rm B}\big).
\label{eq:E}
\end{align}
In going from the third line to the fourth line we applied the gauge transformation $\tau_{ij}\to \tau_{ij}S_iS_j$. We then used Eq.~(\ref{eq:tau_12}), noting that the bond $(12)$ in Eq.~(\ref{eq:tau_12}) is arbitrary and may be replaced by a generic $(ij)$.

Equation~(\ref{eq:E}) shows that the singular part of the energy on the NL is governed by the energy singularity of the pure Ising model at $\beta_e$. In particular, the energy exhibits a nonanalyticity at $\beta_e(\beta)=\beta_c$ (in two and higher dimensions), reflecting the criticality inherited from the disorder ensemble discussed in Sec.~\ref{sec:Probability_distr}. If this nonanalyticity on the NL corresponds to the multicritical point where the paramagnetic, ferromagnetic, and spin-glass phases meet as in the Edwards-Anderson model, the location of the multicritical point of the present model is determined by $\beta_e=\beta_c$.

\subsection{Specific heat}
\label{sec:specific_heat}
We next evaluate the specific heat on the NL:
\begin{align}
&T^{2}C(\beta,\beta)\nonumber\\
&=\frac{1}{A}\sum_{\tau}Z_{\tau}(\beta)e^{\beta\sum \tau_{ij}}
  \left( \frac{\partial_{\beta}^{2} Z_{\tau}(\beta)}{Z_{\tau}(\beta)}
  - \left( \frac{\partial_{\beta}Z_{\tau}(\beta)}{Z_{\tau}(\beta)}
  \right)^{2} \right)\nonumber\\
&=\frac{1}{2^{N}A}\sum_{\tau}Z_{\tau}(\beta)^{2}
  \left( \frac{\partial_{\beta}^{2} Z_{\tau}(\beta)}{Z_{\tau}(\beta)}
  - \left( \frac{\partial_{\beta}Z_{\tau}(\beta)}{Z_{\tau}(\beta)}
  \right)^{2} \right),
\label{eq:C}
\end{align}
where $T=1/\beta$, $N$ is the total number of spins and the gauge transformation $\tau_{ij}\to\tau_{ij}S_iS_j$ has been applied to obtain the last line from the preceding one. Since the intermediate steps are somewhat lengthy, we state here only the results for the first and second terms in Eq.~(\ref{eq:C}) and defer the details to the Appendix.

The first term is
\begin{align}
&\frac{1}{2^{N}A}\sum_{\tau}Z_{\tau}(\beta) \partial_{\beta}^{2}Z_{\tau}(\beta) \notag \\
&= N_{\rm B} + \left(\frac{1}{2}\tanh 2\beta \right)^{2}
   \sum_{\langle ij \rangle \neq \langle kl \rangle}
   \big(1 + \langle S_{i}S_{j} \rangle_{\beta_{e}}^{\rm I} \notag\\
&\hspace{25mm}
   + \langle S_{k}S_{l} \rangle_{\beta_{e}}^{\rm I}
   + \langle S_{i}S_{j}S_{k}S_{l} \rangle_{\beta_{e}}^{\rm I} \big),
\end{align}
and the second term is
\begin{align}
&\frac{1}{2^{N}A}\sum_{\tau}(\partial_{\beta}Z_{\tau}(\beta))^{2} \notag \\
&= \left(\frac{1}{2} \tanh 2\beta \right)^{2}
   \sum_{\langle ij \rangle \neq \langle kl \rangle}
   \big(1 + \langle S_{i}S_{j} \rangle_{\beta_{e}}^{\rm I}
   + \langle S_{k}S_{l} \rangle_{\beta_{e}}^{\rm I}\notag\\
&\hspace{25mm}
   + \langle S_{i}S_{j}S_{k}S_{l} \rangle_{\beta_{e}}^{\rm I} \big)
   + \sum_{\langle ij \rangle} \langle S_{i}S_{j} \rangle_{\beta_{e}}^{\rm I}.
\end{align}
Subtracting the second term from the first, we obtain the exact identity
\begin{equation}
T^{2}C(\beta,\beta) = N_{\rm B}
  - \sum_{\langle ij \rangle} \langle S_{i}S_{j} \rangle_{\beta_{e}}^{\rm I}
  = N_{\rm B} + E_{\rm I}(\beta_{e}).
\label{eq:C_final}
\end{equation}
Thus, on the NL, the specific heat is expressed by a simple function of the energy, not the specific heat, of the pure Ising model at $\beta_e$. Consequently, $C(\beta,\beta)$ stays finite at $\beta_e=\beta_c$, while it inherits the same nonanalyticity as the energy of the pure Ising model.

It is useful to distinguish the temperature derivative of the energy along the NL from the partial derivative at fixed $\beta_p$.
Along the NL, the total derivative reads
\begin{align}
\frac{d}{d\beta}E(\beta,\beta)
=\frac{\partial E(\beta,\beta_p)}{\partial \beta}\Big|_{\beta_p=\beta}
+\frac{\partial E(\beta,\beta_p)}{\partial \beta_p}\Big|_{\beta_p=\beta}.
\label{eq:total_derivative_energy}
\end{align}
The first term on the right-hand side is essentially the specific heat evaluated above and therefore remains finite at $\beta_e=\beta_c$. Since $E(\beta,\beta)$ itself has the Ising-energy singularity through Eq.~(\ref{eq:E}), the total derivative $\frac{d}{d\beta}E(\beta,\beta)$ generally diverges at $\beta_e=\beta_c$ (in two and higher dimensions). We thus conclude that the second term in Eq.~(\ref{eq:total_derivative_energy}), i.e.\ the $\beta_p$-derivative of the energy evaluated on the NL, diverges at $\beta_e=\beta_c$ with the same critical exponent as the temperature derivative of $E_{\rm I}(\beta_e)$.

\section{Gauge non-invariant quantities}
\label{sec:gauge_non_inv}
In the previous section, we derived exact results for gauge-invariant quantities on the NL. We next analyze gauge-noninvariant quantities such as the magnetization and spin correlations. A key observation is that, on the NL, disorder averaging restores an exact correspondence with the pure Ising model for any observable that depends only on the spin variables, not on $\tau$.

\subsection{Reduction to the pure Ising model on the NL}
\label{subsec:reduction}
Let $f(S)$ be an arbitrary function of spins, independent of $\tau$. Then, on the NL, its thermal average followed by the configurational average satisfies
\begin{align}
[\langle f(S)\rangle_{\beta}]_{\beta}
&= \frac{1}{A}\sum_{\tau}Z_{\tau}(\beta)e^{\beta\sum\tau_{ij}}
   \frac{\sum_{S}f(S)\,e^{\beta\sum\tau_{ij}S_{i}S_{j}}}{Z_{\tau}(\beta)}\nonumber\\
&=\frac{1}{A}\sum_{S}f(S)\prod_{\langle i,j\rangle}
  (e^{\beta}e^{\beta S_iS_j}+e^{-\beta}e^{-\beta S_iS_j})\nonumber\\
&=\langle f(S)\rangle_{\beta_{e}}^{\rm I}.
\label{eq:f(S)}
\end{align}
Thus, any $\tau$-independent observable on the NL is exactly equal to the corresponding observable of the pure Ising model at the effective inverse temperature $\beta_e$.

\subsection{Magnetization and its distribution}
As a direct application of Eq.~(\ref{eq:f(S)}), the magnetization on the NL coincides with that of the pure Ising model:
\begin{align}
m(\beta,\beta)=m_{\rm I}(\beta_{e}).
\label{eq:m_equals_mI}
\end{align}
Here we assume symmetry-breaking fixed boundary conditions so that the magnetization does not vanish trivially by symmetry.

A further exact identity, analogous to that in the Edwards-Anderson model~\cite{Nishimori1981,Nishimori2001}, is that the magnetization on the NL equals the spin-glass order parameter:
\begin{align}
&m(\beta,\beta) =
\frac{1}{A}\sum_{\tau}Z_{\tau}(\beta)e^{\beta\sum\tau_{ij}}
  \frac{\sum_S S_i e^{\beta \sum\tau_{ij}S_iS_j}}{Z_{\tau}(\beta)}\notag\\
&=
\frac{1}{2^NA}\sum_{\tau}Z_{\tau}(\beta)\sum_{\sigma}\sigma_i
  e^{\beta\sum \tau_{ij}\sigma_i\sigma_j}
  \frac{\sum_S S_i e^{\beta \sum\tau_{ij}S_iS_j}}{Z_{\tau}(\beta)}
\nonumber \\
&=
\frac{1}{2^NA}\sum_{\tau}Z_{\tau}(\beta)^2
  \Big(\frac{\sum_S S_i e^{\beta \sum\tau_{ij}S_iS_j}}{Z_{\tau}(\beta)}\Big)^2
\nonumber \\
&=q(\beta,\beta).
\label{eq:m_equals_q}
\end{align}
The factor $2^N$ arises from the same gauge-transformation step used in Sec.~\ref{sec:E_and_C}, which converts the weight $Z_{\tau}(\beta)e^{\beta\sum\tau_{ij}}$ into $Z_{\tau}(\beta)^2/2^N$ on the NL. Equation~(\ref{eq:m_equals_q}) implies that there is no spin-glass phase on the NL in the usual sense ($m=0$ with $q>0$), as in the Edwards-Anderson model.

More generally, Eq.~(\ref{eq:f(S)}) implies that the distribution of the magnetization on the NL is identical to that of the pure Ising model at $\beta_e$:
\begin{align}
&P(m,\beta,\beta)\nonumber\\
&=\frac{1}{A}\sum_{\tau}Z_{\tau}(\beta)e^{\beta\sum\tau_{ij}}
  \frac{\sum_{S}\delta\big(m-\frac{1}{N}\sum_i S_i\big)
  \,e^{\beta\sum\tau_{ij}S_{i}S_{j}}}{Z_{\tau}(\beta)}
\nonumber\\
&=P_{\rm I}(m,\beta_e).
\label{eq:P_m_equals_P_I}
\end{align}
A similar manipulation shows that the spin-glass order-parameter distribution on the NL is the same:
\begin{align}
P(q,\beta,\beta)=P_{\rm I}(m,\beta_e).
\label{eq:P_q_equals_P_I}
\end{align}
Since the pure Ising magnetization distribution $P_{\rm I}(m,\beta_e)$ has at most two delta-function peaks in the thermodynamic limit (corresponding to $m=\pm m_0$ in the ordered phase, and a single peak at $m=0$ in the disordered phase), Eqs.~(\ref{eq:P_m_equals_P_I}) and (\ref{eq:P_q_equals_P_I}) imply that there is no replica symmetry breaking on the NL. This contrasts with the behavior found for the correlated-disorder distribution $P_1(\tau)$ in Eq.~(\ref{eq:prob_dist_old1})~\cite{Nishimori2024}. Thus, the existence or absence of replica symmetry breaking on the NL depends on the type of correlations in disorder.

\subsection{Correlation function and critical exponents on the NL}
The two-point correlation function obeys the general identity from Eq.~(\ref{eq:f(S)}),
\begin{align}
\big[ \langle S_i S_j \rangle_{\beta}\big]_{\beta}
= \langle S_i S_j \rangle_{\beta_{e}}^{\rm I}.
\label{eq:correlation}
\end{align}
Consequently, the critical exponents $\nu$ and $\eta$ at the critical point reached along the NL coincide with the corresponding exponents of the pure Ising model.

The magnetic susceptibility is defined as
\begin{align}
\chi(\beta,\beta)
=\sum_{i,j}\big[ \langle S_i S_{j}\rangle_{\beta}\big]_{\beta}
  -\big[\big(\sum_{i}\langle S_i \rangle_{\beta}\big)^{2}\big]_{\beta}.
\label{eq:susceptibility}
\end{align}
The first term is exactly the pure-Ising correlation sum by Eq.~(\ref{eq:correlation}), whereas the second term is not straightforward to evaluate in a closed form. Nevertheless, using Eq.~(\ref{eq:f(S)}) we can obtain the upper bound as
\begin{align}
\chi(\beta,\beta)
\le \sum_{i,j}\big[ \langle S_i S_{j}\rangle_{\beta}\big]_{\beta}
  -\big[\sum_{i}\langle S_i \rangle_{\beta}\big]_{\beta}^{2}
\label{eq:suscep_bound}
\end{align}
In the paramagnetic phase, where $\langle S_i\rangle_{\beta}=0$, the second term of Eqs.~(\ref{eq:susceptibility}) and (\ref{eq:suscep_bound}) vanishes, and the divergence of $\chi(\beta,\beta)$ upon approaching the critical point along the NL from the paramagnetic side is governed by the same critical exponent $\gamma$ as in the pure Ising model.

\subsection{An inequality and the location of the NL intersection}
\label{sec:inequality}
It is possible to derive the following inequality in exactly the same manner as for the Edwards-Anderson model~\cite{Nishimori1981,Nishimori2001}, that is, by applying the gauge transformation to the left-hand side of the following equation to obtain the middle expression,
\begin{align}
\big|[\langle S_i\rangle_{\beta}]_{\beta_p}\big|
= \big|[\langle S_i\rangle_{\beta}\langle S_i\rangle_{\beta_p}]_{\beta_p}\big|
\le [\big|\langle S_i\rangle_{\beta_p}\big|]_{\beta_p}.
\label{eq:ineq_m}
\end{align}
The right-hand side is a trivial consequence of the middle expression after moving the absolute value $|\cdots |$ from the outside to inside the configurational average $[\cdots]_{\beta_p}$.
If the system is paramagnetic on the NL (i.e., if the right-hand side vanishes), Eq.~(\ref{eq:ineq_m}) implies that the magnetization vanishes at any temperature $\beta$ (the left-hand side) with the same $\beta_p$. Therefore there is no ferromagnetic phase below the crossing point of the NL and the ferromagnetic phase boundary. As a consequence, the boundary between the ferromagnetic and non-ferromagnetic (spin glass or paramagnetic) phases is either vertical or reentrant, strongly suggesting that this crossing point is the multicritical point, as in the Edwards-Anderson model \cite{Honecker2001,LeDoussal1988,Hasenbusch2007,Hasenbusch2008,Ceccarelli2011,Katzgraber2006}.

\section{Gaussian distribution}
\label{sec:Gaussian}
So far we have analyzed the case of discrete couplings $\tau_{ij}=\pm 1$.
A continuous Gaussian version can be treated in essentially the same way, because the disorder distribution is constructed from the partition function of the same spin system and the gauge transformation on the NL works identically. We therefore state the definitions and results without repeating the intermediate steps.

We consider a model with real-valued couplings $J_{ij}$ and define the correlated-disorder distribution by
\begin{equation}
P(J) = \frac{1}{A}\, Z_{J}(\beta_{p})
\exp\!\Bigg(-\frac{1}{2}\sum_{\langle ij\rangle} J_{ij}^{2} + \beta_{p}\sum_{\langle ij\rangle} J_{ij}\Bigg),
\label{eq:gauss_P}
\end{equation}
where
\begin{equation}
Z_{J}(\beta_{p})=\sum_{S}\exp\!\Bigg(\beta_{p}\sum_{\langle ij\rangle} J_{ij}S_iS_j\Bigg)
\end{equation}
and $A$ is the normalization factor.

The normalization factor is obtained by completing the square in each Gaussian integral:
\begin{align}
A &= \int \prod_{\langle ij \rangle} \frac{dJ_{ij}}{\sqrt{2\pi}}
       Z_{J}(\beta_{p})
       \exp\!\Bigg(-\frac{1}{2}\sum_{\langle ij\rangle} J_{ij}^{2} + \beta_{p}\sum_{\langle ij\rangle} J_{ij}\Bigg) \nonumber\\
  &= e^{N_{\rm B} \beta_p^2}\, Z_{\mathrm{I}}(\beta_{e}),
\label{eq:gauss_A}
\end{align}
where $Z_{\mathrm{I}}(\beta_{e})$ is the partition function of the pure Ising model and
\begin{equation}
\beta_{e}(\beta_p) = \beta_p^2
\label{eq:gauss_beta_e}
\end{equation}
is the effective inverse temperature (the Gaussian analog of Eq.~(\ref{eq:effective_beta})).

On the NL $\beta_p=\beta$, the energy is
\begin{align}
E(\beta, \beta)
= \beta\big( E_{\mathrm{I}}(\beta_{e})-N_{\rm B} \big),
\label{eq:gauss_E}
\end{align}
and the specific heat satisfies the same relation as in the discrete case:
\begin{equation}
T^{2} C(\beta, \beta) = N_{\rm B} + E_{\mathrm{I}}(\beta_{e}).
\label{eq:gauss_C}
\end{equation}
Thus, the discussion at the end of Sec.~\ref{sec:specific_heat} applies to the Gaussian model as well.

All the identities and inequalities derived in Sec.~\ref{sec:gauge_non_inv} carry over to the Gaussian case without modification, because they rely only on the NL condition and the gauge-transformation structure.

\section{Summary and discussion}
\label{sec:summary}

We have defined an Ising spin-glass model with correlations among the disorder variables, controlled by the parameter $\beta_p$ in the disorder distribution~(\ref{eq:prob_dist}). The disorder correlations are short-ranged except at the point where the effective inverse temperature $\beta_e(\beta_p)$ coincides with the critical inverse temperature of the pure Ising model, $\beta_e=\beta_c$.  In particular, the disorder distribution itself becomes singular at $\beta_e=\beta_c$, reflecting the criticality of the pure Ising model that enters the normalization factor.

On the Nishimori line (NL), $\beta_p=\beta$, we proved a set of exact identities which reduce many observables of the present correlated-disorder model to those of the pure Ising ferromagnet at the effective inverse temperature $\beta_e$.  The internal energy on the NL is expressed by the energy of the pure Ising model up to a constant and a trivial prefactor, and, more strikingly, the specific heat on the NL is expressed by the energy (not the specific heat) of the pure Ising model, again up to a constant and a trivial factor.  Consequently, the specific heat stays finite along the NL, including at $\beta_e=\beta_c$, while the singular parts of the energy and the specific heat on the NL are governed by the singularity of the pure Ising energy.

We also showed that gauge-noninvariant quantities on the NL, including the magnetization and correlation functions, coincide exactly with the corresponding quantities of the pure Ising ferromagnet at $\beta_e$.  Therefore, if the NL critical point is indeed the multicritical point  where the paramagnetic, ferromagnetic, and spin-glass phases meet, the leading critical behavior at this multicritical point is pure-Ising-like. For example, the exponents $\nu$, $\eta$, and $\gamma$ on the paramagnetic side are identical to those of the pure Ising model.  This should be contrasted with the conventional universality class associated with the multicritical point of the Edwards-Anderson model, which has been studied extensively by renormalization-group and numerical methods~\cite{Honecker2001,LeDoussal1988,Hasenbusch2007,Hasenbusch2008,ParisenToldin2010}.  In this sense, the present model suggests a different multicritical universality class than the conventional universality class in the Edwards-Anderson model without disorder correlation at the multicritical point, characterized by the pure-Ising criticality.

We note that it has not been rigorously proven that the critical point on the NL of the present model is the multicritical point.  Nevertheless, for the Edwards-Anderson model with independent disorder, it is generally believed that the NL critical point coincides with the multicritical point based on numerical and renormalization-group analyses~\cite{LeDoussal1988,Hasenbusch2008,Ceccarelli2011,Katzgraber2006}, and we expect an analogous scenario here partly based on the inequality derived in Sec.~\ref{sec:inequality}.  Under this expectation, our exact results provide an analytically controlled example in which the multicritical point exhibits pure-Ising criticality and in which the thermodynamic singularities on the NL can be determined explicitly.

The idea of the present model has been inspired by our previous correlated-disorder constructions~\cite{Nishimori2024,Nishimori2025,Nishimori2025tc}, and both can be seen as manifestations of the replica formalism with different replica numbers as discussed in Sec.~\ref{sec:replica}. Nevertheless, in contrast to those previous models, (i) the present model interpolates smoothly between the pure Ising model ($\beta_p\to\infty$) and the Edwards-Anderson model with symmetric independent disorder ($\beta_p=0$), and (ii) it admits an exact reduction of many observables on the NL to the corresponding quantities of the pure Ising model, enabling an explicit discussion of the critical behavior at $\beta_e=\beta_c$.

The present results raise several questions about the role of correlations in disorder.  For example, what critical properties does the para-ferro boundary have between the pure Ising limit ($\beta_p\to\infty$) and the NL ($\beta_p=\beta$)?  This issue is closely related to the applicability of the Harris criterion~\cite{Harris1974}: While the latter criterion addresses the instability of pure criticality against independent weak disorder, it is not obvious how it extends to correlated disorder of the present type, especially given that the NL critical behavior is exactly tied to that of the pure Ising model in the present type of disorder.

The problem of correlated disorder may also be relevant to quantum information processing, where correlations in noise can be important and where spin-glass theory has been used to analyze error thresholds in the absence of correlations in noise ~\cite{Dennis2002,Wang2003}. 

We point out that that the present and previous constructions \cite{Nishimori2024,Nishimori2025,Nishimori2025tc} may serve as a step toward systematic analyses of correlations in disorder variables, a topic that has largely escaped extensive studies. In particular, we hope that the present exactly tractable setting on the NL will be useful as a reference point for systematic studies of how correlations in disorder modify critical phenomena in magnetic systems, quantum error correction, and beyond.

\section*{Data availability}
No data were created or analyzed in this study.

\section*{Appendix}
This Appendix shows the detailed derivation of the specific heat as discussed in Sec.~\ref{sec:specific_heat},
\begin{align}
&T^{2}C(\beta,\beta)\nonumber\\
&=\frac{1}{2^{N}A}\sum_{\tau}Z_{\tau}(\beta)^{2} \left( \frac{\partial_{\beta}^{2} Z_{\tau}(\beta)}{Z_{\tau}(\beta)} - \left( \frac{\partial_{\beta}Z_{\tau}(\beta)}{Z_{\tau}(\beta)} \right)^{2} \right).\label{eq:Ca}
\end{align}
The first term is evaluated as
\begin{widetext}
\begin{align}
&\frac{1}{2^{N}A}\sum_{\tau}Z_{\tau}(\beta) \partial_{\beta}^{2}Z_{\tau}(\beta) \notag \\
&= \frac{1}{2^{N}f_{1}(\beta)Z_{\rm I}(\beta_{e})} \sum_{\tau} \sum_{S,\sigma} e^{\beta \sum \tau_{ij}S_{i}S_{j}} \sum_{\langle ij \rangle} \sum_{\langle kl \rangle} \tau_{ij}\tau_{kl}\sigma_{i}\sigma_{j}\sigma_{k}\sigma_{l} e^{\beta \sum \tau_{ij}\sigma_{i}\sigma_{j}} \notag \\
&= \frac{1}{2^{N}f_{1}(\beta)Z_{\rm I}(\beta_{e})} \left\{ \sum_{\langle ij \rangle \neq \langle kl \rangle} \sum_{S,\sigma} \sum_{\tau} \tau_{ij}\tau_{kl} \sigma_{i}\sigma_{j}\sigma_{k}\sigma_{l} e^{\beta \sum \tau_{ij}S_{i}S_{j} + \beta \sum \tau_{ij}\sigma_{i}\sigma_{j}} \right. \notag \\
&\quad \left. + \sum_{\langle ij \rangle} \sum_{S,\sigma} \sum_{\tau} e^{\beta \sum \tau_{ij}S_{i}S_{j} + \beta \sum \tau_{ij}\sigma_{i}\sigma_{j}} \right\} \notag \\
&= \frac{1}{2^{N}f_{1}(\beta)Z_{\rm I}(\beta_{e})} \left\{ \sum_{\langle ij \rangle \neq \langle kl \rangle} \sum_{S,\sigma} \sum_{\tau} \tau_{ij}\tau_{kl} \sigma_{i}\sigma_{j}\sigma_{k}\sigma_{l} e^{\beta \sum \tau_{ij}S_{i}S_{j} + \beta \sum \tau_{ij}\sigma_{i}\sigma_{j}} \right. \notag \\
&\quad \left. + N_{\rm B} \cdot 2^{N}f_{1}(\beta)Z_{\rm I}(\beta_{e}) \right\} \notag \\
&= N_{\rm B} + \frac{1}{2^{N}f_{1}(\beta)Z_{\rm I}(\beta_{e})} \sum_{\langle ij \rangle \neq \langle kl \rangle} \sum_{S} f_{1}(\beta) \frac{(\sinh 2\beta)^{2}}{(2 \cosh 2\beta)^{2}} \sum_{\sigma} e^{\beta_{e}\sum S_{i}S_{j}} (1+S_{i}S_{j})(1+S_{k}S_{l}) \notag \\
&= N_{\rm B} + \left(\frac{1}{2}\tanh 2\beta \right)^{2} \sum_{\langle ij \rangle \neq \langle kl \rangle} (1 + \langle S_{i}S_{j} \rangle_{\beta_{e}}^{\rm I} + \langle S_{k}S_{l} \rangle_{\beta_{e}}^{\rm I} + \langle S_{i}S_{j}S_{k}S_{l} \rangle_{\beta_{e}}^{\rm I} ).
\end{align}
The second term is
\begin{align}
&\frac{1}{2^{N}A}\sum_{\tau}(\partial_{\beta}Z_{\tau}(\beta))^{2} \notag \\
&= \frac{1}{2^{N}f_{1}(\beta)Z_{\rm I}(\beta_{e})} \left\{ \sum_{\langle ij \rangle \neq \langle kl \rangle} \sum_{S,\sigma} S_{i}S_{j}\sigma_{k}\sigma_{l} \sum_{\tau} \tau_{ij}\tau_{kl} e^{\beta \sum \tau_{ij}S_{i}S_{j} + \beta \sum \tau_{ij}\sigma_{i}\sigma_{j}} \right. \notag \\
&\quad \left. + \sum_{\langle ij \rangle} \sum_{S,\sigma} S_{i}S_{j}\sigma_{i}\sigma_{j} \sum_{\tau} e^{\beta \sum \tau_{ij}S_{i}S_{j} + \beta \sum \tau_{ij}\sigma_{i}\sigma_{j}} \right\} \notag \\
&= \frac{1}{Z_{\rm I}(\beta_{e})} \sum_{\langle ij \rangle \neq \langle kl \rangle} \sum_{S} S_{i}S_{j} e^{\beta_{e} \sum S_{i}S_{j}} \frac{(1+S_{i}S_{j})(1+S_{k}S_{l}) \sinh^{2} 2\beta}{(2 \cosh 2\beta)^{2}} + \sum_{\langle ij \rangle} \langle S_{i}S_{j} \rangle_{\beta_e}^{\rm I} \notag \\
&= \left(\frac{1}{2} \tanh 2\beta \right)^{2} \sum_{\langle ij \rangle \neq \langle kl \rangle} (1 + \langle S_{i}S_{j} \rangle_{\beta_e}^{\rm I} + \langle S_{k}S_{l} \rangle_{\beta_e}^{\rm I} + \langle S_{i}S_{j}S_{k}S_{l} \rangle_{\beta_e}^{\rm I} ) + \sum_{\langle ij \rangle} \langle S_{i}S_{j} \rangle_{\beta_e}^{\rm I}.
\end{align}
Therefore the total specific heat is
\begin{equation}
T^{2}C(\beta,\beta) = N_{\rm B} - \sum_{\langle ij \rangle} \langle S_{i}S_{j} \rangle_{\beta_e}^{\rm I} = N_{\rm B} + E_{\rm I}(\beta_{e}).
\end{equation}
\end{widetext}

\end{document}